\documentstyle[twocolumn,aps,epsf,floats]{revtex}       
\begin{document}

\twocolumn[\hsize\textwidth\columnwidth\hsize\csname %
@twocolumnfalse\endcsname

\title
{The dispersion of a single hole in an antiferromagnet}
\author{Andrey V. Chubukov and Dirk K. Morr}
\address
{Department of Physics, University of Wisconsin-Madison,
1150 University Ave., Madison, WI 53706}
\date{\today}
\maketitle
\begin{abstract}
We revisit the problem of the
dispersion of a single hole injected into a quantum antiferromagnet. We 
applied a spin-density-wave formalism extended to large 
number of orbitals
and obtained an integral equation for the
full quasiparticle Green's function in the self-consistent ``non-crossing" 
Born approximation. 
We  found that  for $t/J \gg 1$, 
 the bare fermionic dispersion is completely overshadowed by the self-energy
corrections. In this case, the quasiparticle Green's function contains a
broad incoherent continuum which extends over a frequency range of $\sim 6t$. In addition, there exists a narrow region
of width $O(JS)$ below the top of the valence band, where the excitations
are mostly coherent, though with a small quasiparticle residue $Z \sim
J/t$. The top of the valence band is located at $(\pi/2,\pi/2)$.
We found that the form of the fermionic dispersion, and, in particular, the
ratio of the effective masses near $(\pi/2,\pi/2)$ strongly depend on
the assumptions one makes for the form of the magnon propagator. 
We argue in this paper that two-magnon Raman 
scattering as well as neutron scattering experiments
strongly suggest that the zone boundary magnons are not free particles since a substantial portion of
their spectral weight is transferred into an incoherent background.
We modeled this effect by introducing a cutoff $q_c$ in the integration over
 magnon momenta. 
We found analytically that for small $q_c$, 
the strong coupling
solution for the Green's function is universal, and both
effective masses are equal to $(4JS)^{-1}$.
We further computed the full fermionic
dispersion for $J/t =0.4$ relevant for $Sr_2CuO_2Cl_2$, 
 and $t^\prime=-0.4J$ and
found not only that the masses are both equal to $(2J)^{-1}$, but also that
the energies at $(0,0)$ and $(0,\pi)$ are 
equal, the energy at $(0,\pi/2)$ is about half of that at $(0,0)$,
and the bandwidth for the coherent excitations is around $3J$.
All of these results are  in full agreement with the experimental data.
Finally, we found that weakly damped excitations only exist
in a narrow range around $(\pi/2,\pi/2)$.
Away from the vicinity of $(\pi/2,\pi/2)$, the excitations are overdamped,
and the spectral function possesses a broad maximum rather than a 
sharp quasiparticle peak.
This last feature was also reported in photoemission experiments. 
\end{abstract}
\pacs{PACS: } 

]

\narrowtext

\section{Introduction}
The  dispersion of a single hole in a quantum antiferromagnet
is one of the  issues in the field of high-temperature superconductivity
which has attracted a substantial amount of interest over a number of 
years~
\cite{SS,KaneLee,Marsi,SWZ,t-J,Dag,Dag1,Dag2,S-T,Vig,Manous,Trug,Sach,Bel,CM1,Duf,Mukhin,Gooding,Sushkov}. 
The parent compounds of the 
high-$T_c$ materials are quantum Heisenberg antiferromagnets as 
was demonstrated by neutron scattering~\cite{Hayden}, NMR~\cite{Imai}
 and Raman~\cite{Blum} experiments. 
The antiferromagnetic spin ordering strongly modifies the electronic
dispersion which by all accounts is very different from what one would
expect from band theory calculations. 
Upon hole doping, short-range antiferromagnetism gradually disappears, and the 
overdoped cuprates possess an electronic dispersion which 
is consistent with band theory
predictions~\cite{photo}. How the electronic spectrum evolves 
with doping is currently
a subject of intensive experimental and theoretical
studies~\cite{Shen-Schr,Drusha,Morr,Joerg}.  As an important input 
for these
studies, one needs to know what happens in the limit of zero doping
when a single hole is injected into
a quantum antiferromagnet.  
 
The dispersion of a single hole in an antiferromagnet has been 
intensively studied experimentally  and theoretically . 
Experimental information comes
 from photoemission experiments on the half-filled
$Sr_2CuO_2Cl_2$ which is not a high-$T_c$ superconductor, but contains
the same $CuO_2$ planes as the high-$T_c$ materials~\cite{Wells,LaRosa}.
Most of the theoretical analysis was 
performed in the framework of the $t-J$ and Hubbard models which are widely 
believed to adequately describe the low-energy physics of the underlying 
three-band model~
\cite{SS,KaneLee,Marsi,SWZ,t-J,Dag,Dag1,Dag2,S-T,Vig,Manous,Trug,Sach,CM1,Duf,Mukhin,Gooding,Sushkov}.  Early analytical and
numerical computations were performed in the antiferromagnetically ordered phase and for the  case when a hopping is only possible between nearest neighbors~\cite{SS,KaneLee,Marsi,t-J,Trug,Sach}. 
These studies have shown that in the strong coupling limit (large $U$
limit in the Hubbard model or $t \gg J$ limit in the $t-J$ model), the
Green's function of a single hole has the form 
\begin{equation}
G (k, \omega) = \frac{Z}{\omega - E_k} + G_{inc} (k,\omega) \ ,
\end{equation}
where the coherent part is confined to scales smaller than $2J$, while the
incoherent background stretches upto a few $t$. The quasi-particle residue 
of the coherent piece is small and scales as $Z \propto J/t$ in the 
limit $t \gg J$. The dispersion $E_k$ has a maximum at $k = (\pi/2,
\pi/2)$ and symmetry related points. All calculations have demonstrated that
the dispersion around this point is very anisotropic with a substantially
larger mass along the
$(0,\pi)$ to $(\pi,0)$ direction than along the Brillouin zone diagonal.
For $t/J =2.5$ relevant to cuprates,
the ratio of the masses is about $5-7$ in the $t-J$ model 
(without a three-site term)~\cite{Marsi}, 
and it is even larger in the Hubbard  model due to the presence
of the bare dispersion $J(cos k_x + cos k_y)^2$ which yields an extra contribution to  the mass along the zone diagonal~\cite{Zaanen}.
 
It turns out, however, that the experimental results for $Sr_2CuO_2Cl_2$ 
\cite{LaRosa,Wells} are rather different from these predictions. 
Although the photoemission data
have demonstrated that the maximum of $E_k$ is at 
$k =(\pi/2,\pi/2)$ consistent with the theory, the experimentally measured
ratio of the masses is close to one in clear disagreement with the
theoretical predictions.
Moreover, the data show that the coherent peak in the spectral function
 exists only in
a narrow region around $(\pi/2,\pi/2)$ while away from this region, the 
hole spectral function is nearly featureless. This implies that the
fermionic excitations become overdamped already at 
energies which are substantially smaller than $2J$. 

After the data were reported, several attempts have been made 
to improve the
agreement between theory and experiment. One scenario was 
 put forward by researchers working on the ``gauge theory" approach 
to cuprates~\cite{gauge}, most recently by
Laughlin~\cite{Laughlin}. 
He  argued that the isotropy of the dispersion
together with the observed mostly
incoherent nature of the electronic excitations are 
signatures of a spin-charge separation.  For a 
state where spin and charge degrees of freedom  are
described by separate quasiparticles (spinons and holons, respectively),
 the electron Green's function is just a convolution of the spinon and
holon propagators. It does not have a pole which normally would be
associated with the coherent part of $G(k,\omega)$, but rather a branch cut
which describes fully incoherent excitations. Laughlin argued that 
since spinon and holon energies are well separated (the spinon energy
has an overall scale of $J$, while the holon energy is $O(t)$),
the position of the branch cut virtually 
coincides with the spinon dispersion.
 In the mean-field theory for the spin-charge separated state,
 the spinon energy has the form
\begin{equation}
E^{spinon}_ k = - C_{sw}(\cos^2{k_x}+\cos^2{k_y})^{1/2} \ ,
\end{equation}
where $C_{sw} \sim 1.6J$ is the spin-wave velocity in a 2D $S=1/2$ 
antiferromagnet.  This dispersion has an 
isotropic maximum at $k = (\pi/2,\pi/2)$, a bandwidth of $2.2J$
and equal energies for $k=(0,0)$ and $(0,\pi)$ - all of these features
are consistent with the data together with the near absence of the
quasiparticle peak.

An obvious weakness of the mean-field analysis
of spinons and holons is that it neglects 
the effects due to a gauge field. Beyond the mean-field level,
a gauge field may glue spinons and holons into a bound state thus rendering
the electron as a coherent quasiparticle. 
Laughlin conjectured that the confinement 
takes place only below $T_N$, while
the experimental data were actually collected at 
$T=350K$ which is $100K$ above 
the Neel temperature.  He then proposed that
if measurements are done at much lower temperatures, they should yield
an anisotropic dispersion consistent with the results 
obtained in the ordered state with no spin-charge separation.

Another, more conventional approach to the single hole problem
assumes that there is no spin-charge separation at any $T$, and that
the experimental data in fact reflect the behavior of the hole dispersion
in the antiferromagnetically ordered phase~
\cite{CM1,Duf,Mukhin,Gooding,Sushkov}.
 Within this approach, 
the discrepancy with the
data is mainly 
attributed to the fact that the original model did not contain
a hopping term $t^{\prime}$ between next-nearest neighbors (and, 
possibly, also between further neighbors). The presence of the 
a finite $t^{\prime}$ term in the Hubbard model is justified, 
at least partly,
by studies which derived an effective one-band model from the underlying
three-band model by comparing the energy levels around the charge transfer
gap \cite{Hyber}. These studies predicted that the 
second-neighbor hopping is about 
$t^{\prime} = -0.2 t$. By itself, this hopping is small
compared to $t$. 
However, in an antiferromagnetic background, the hole can only move 
within the same sublattice, otherwise the antiferromagnetic ordering is 
disturbed. The hopping term $t^\prime$  connects the sites 
within the same sublattice, and therefore is not 
affected by antiferromagnetism.
On the contrary, the $t$ term contributes to the hopping within a sublattice
only via the creation of a virtual doubly occupied state which
costs the energy $U$.  As a result, the $t-$part of the dispersion 
is rescaled and becomes of order $t^2/U = O(J)$.  
One therefore has to compare $t^{\prime}$ not with $t$ but rather 
with $J$. For  $J/t \sim 0.4$, we then obtain $t^{\prime} =-0.5J$, which 
immediately implies that the corrections due to $t^\prime$ are actually 
quite relevant. 

It has been mentioned several times in the literature that
the inclusion of $t^{\prime} =-0.5J$ 
into the Hubbard model yields a good agreement with the experimental data
already at the mean-field level~\cite{CM1,Duf}.
 Indeed, the mean-field spin-density-wave
(SDW) formula for the hole dispersion at large $U$ is
\begin{equation}
E_ k = -J(\cos{k_x}+\cos{k_y})^2 -4t^\prime \cos{k_x} \cos{k_y} \ .
\label{a}
\end{equation}
For $t^{\prime} = -0.5J$, this formula transforms into
\begin{equation}
E_k = -J(\cos^2{k_x}+\cos^2{k_y})
\end{equation}
(here we assumed that the chemical potential is at the 
top of the valence band).
This dispersion possesses two equal effective masses if one expands
around the maximum at $(\pi/2,\pi/2)$, and has a 
a local maximum at $(0,\pi/2)$ with $E =-J$. Both of these 
results are consistent with the most recent data by LaRosa {\it et
al.}~\cite{LaRosa}. 
Furthermore, the data show that the energies at $(0,0)$ and
$(0,\pi)$ are both equal to $-2J$. This also
agrees with the photoemission data~\cite{Wells,LaRosa}.

The conventional mean-field SDW-type approach also possesses the weakness that it predicts fully coherent excitations upto  $2J$. 
The data, however,  demonstrate that
away from the vicinity of $(\pi/2,\pi/2)$,
the coherent part of the dispersion is almost completely
overshadowed by the incoherent background.  
Earlier studies \cite{CM1} which went beyond 
the mean-field level have demonstrated that self-energy corrections reduce
the quasiparticle residue thus transferring part of the spectral weight into
the incoherent background. However, 
these corrections also
effectively decrease $t^{\prime}$ and thus render the spectrum more 
anisotropic (see Fig.~\ref{edisp6} and \ref{edisp4} below).
 From this perspective,
the observed isotropy of 
the dispersion around $(\pi/2,\pi/2)$ is attributed  in a conventional approach
to some fine tuning of both $J/t$ and 
$t^\prime/J$ and is therefore completely accidental~\cite{Oleg}.

In this paper we show 
 that in a certain limit specified below, 
the near-degeneracy of the spectrum around $(\pi/2,\pi/2)$ turns out to be 
a fundamental, universal property of a single hole in an antiferromagnet, 
independent of the details of the physics at atomic scales.   
Our key point is this: in all previous studies which yielded
anisotropic spectra, it was assumed that magnons behave as free particles
for all momenta.
In this case, the integral over the magnon momenta in the self-energy 
term runs over the whole  magnetic Brillouin zone (MBZ). 
On the other hand, Raman studies
of the two-magnon profile in the insulating parent compounds
of high $T_c$ materials
 have demonstrated  that the width of the two-magnon peak is much
broader than one would expect for free magnons~\cite{Blum,Sin90}. 
The dominant contribution to
this peak comes from the magnons near the boundary of the MBZ. 
Complimenting these findings, neutron scattering experiments on
$La_2CuO_4$~\cite{gabe} 
have shown that about half
of the spectral weight of the quasiparticle peak for the 
zone boundary magnons is 
transferred into a broad incoherent background.

It has been suggested that the broadening is due to the strong interaction
between these magnons and phonons~\cite{WebFord,Knoll}. This interaction is
finite and not necessary small at $T=0$ contrary
to the magnon-magnon interaction which gives rise to 
an incoherent part of the magnon spectral function only 
at finite $T$ and is irrelevant for $T \ll J$ \cite{halperin}. 
  
In this situation, it seems reasonable to assume that the
contribution from the zone boundary magnons to the electronic self-energy 
is substantially  reduced compared to what one would obtain for free spin waves.
This however is true only for zone-boundary magnons. 
For long-wavelength magnons, the magnon-phonon vertex scales linearly with
the magnon momentum, and the incoherent part of the magnon propagator
is small. 
The simplest way to model this effect is to  introduce an upper cutoff 
$q_c$ in the integration over magnon momenta. Naively, one might expect that
the hole dispersion would strongly depend on $q_c$. 
However, we will
demonstrate that at large $t/J$, when the bare dispersion is
irrelevant, only the quasiparticle residue does depend on $q_c$, while the
effective masses are in fact independent of $q_c$ in the limit when $q_c$ is
sufficiently small. We  explicitly show that
in this limit, both masses turn out to be equal to $1/2J$.
The dispersion near $(\pi/2,\pi/2)$ is then isotropic and has a form 
$E_k =-J {\tilde k}^2$ where ${\tilde k}$ is the deviation from $(\pi/2,\pi/2)$.
Furthermore,  we  show that for a certain range of $q_c$ 
the inclusion of
$t^{\prime} =-0.5J$ extends the region where the two masses are 
approximately equal to basically all values of $t/J$.
This last result allows us to 
correctly reproduce the measured hole dispersion in $Sr_2CuO_2Cl_2$. 

The paper is organized as follows. In the next section, we outline the
formalism and derive the integral equation for the quasiparticle Green's
function by expanding around the mean-field SDW solution. 
In Sec.~\ref{anal_res} we present our analytical results in the large $t/J$
 limit. In this section
we also discuss the role of the vertex corrections to the spin-fermion vertex.
In Sec.~\ref{num_res}, we present the results of the numerical solution of the
self-consistency
equation for the quasiparticle Green's function for different values of $J/t$.
Sec.~\ref{concl} contains a summary of our results.

\section{The Formalism}
\label{form}
As mentioned in the introduction, our starting point for the description of 
the insulating parent compounds of the high-$T_c$ materials is the effective 2D
one-band Hubbard model \cite{SchrAnd,Scal,Kampf}, given by
\begin{equation}
{\cal H} = - \sum_{i,j} t_{i,j} c^{\dagger}_{i,\alpha} c_{j,\alpha} 
+ U \sum_i c^{\dagger}_{i,\uparrow} c_{i,\uparrow} c^{\dagger}_{i,\downarrow} 
c_{i,\downarrow} \ .
\label{hub}
\end{equation}
Here $\alpha$ is the spin index and $t_{i,j}$ is the hopping
integral which we assume to act between nearest and next-nearest neighbors
($t$ and $t^{\prime}$, respectively). Throughout the paper we assume
 that the ground state  of the Hubbard model
is antiferromagnetically ordered. In this situation,
a  way to calculate the spectral function in a systematic perturbative
expansion
is to extend the Hubbard model to a large number
of orbitals, $n=2S$, and use a $1/S$
expansion around the  mean-field SDW state \cite{affleck}.
The $1/S$ expansion for the Hubbard model has been discussed several
 times in the literature~\cite{CM1,Morr}
 and we will use it here without further
clarification.
To obtain the mean-field solution, 
 one introduces an antiferromagnetic long range order  
parameter $S_z = \langle c^{\dagger}_k c_{k+Q} \rangle $ 
and uses it
to decouple the interaction term in Eq.(\ref{hub}). Diagonalizing then the
quadratic form by means of a unitary transformation
one obtains two electronic bands for the conduction and valence fermions, 
whose energy dispersion is given by
\begin{equation}
E^{c,v}_ k = \pm \sqrt{(\epsilon^{-}_ k)^2 + \Delta^2}+
 \epsilon^{+}_k \ ,
\end{equation}
where 
$$
\epsilon^{\pm}_k = { \epsilon_k \pm \epsilon_{k+Q} 
\over 2 } \qquad \Delta = U \langle S_z \rangle
$$
\begin{equation}
\epsilon_ k =-4{\bar t}S(\cos k_x + \cos k_y) - 8{\bar t}^{\prime} S \cos k_x \cos k_y - \mu \ .
\end{equation}
Here $E^{c,v}_ k$ is the dispersion of
the conduction and valence fermions, 
respectively, $\epsilon_k$ is the dispersion of free fermions, 
 $\mu$ is the chemical potential, and $\langle S_z \rangle$ is the sublattice
magnetization. To facilitate the $1/S$ expansion,
we also introduced ${\bar t} = t/2S$ and ${\bar t}^{\prime} = t^\prime /2S$.
In the large-$U$ limit which we only consider,  one
can expand the square root and obtains
\begin{eqnarray}
E^{c,v}_ k &=& \pm \Delta \pm 2JS(\cos{k_x}+\cos{k_y})^2 \nonumber \\
& & \quad  -8{\bar t}^\prime S\cos{k_x} \cos{k_y} - \mu \ ,
\end{eqnarray}
where $J = 4{\bar t}^2/U$. At half-filling, the chemical potential can be set 
to the top of the valence band $(\mu = -\Delta)$; for $S=1/2$ we then
reproduce Eq.(\ref{a}) from the introduction.

At infinite $S$, the mean-field approach is exact. At finite $S$,
the bare Green's function is renormalized due to 
the interaction with spin waves. 
\begin{figure} [t]
\begin{center}
\leavevmode
\epsfxsize=7.5cm
\epsffile{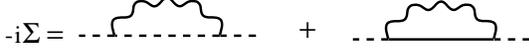}
\end{center}
\caption{ The lowest order self-energy correction for the valence 
fermions in the SDW model. The solid and dashed lines are
the bare propagators of conduction and 
valence fermions, respectively. The wavy line describes transverse spin 
fluctuations.}
\label{sigma}
\end{figure}
The lowest order self-energy corrections for 
valence fermions are given by the diagrams in Fig.~\ref{sigma}.
The solid and dashed lines in these diagrams are the propagators of
 conduction and valence
fermions, respectively. The wavy lines describe transverse 
spin fluctuations which in the SDW
approach are collective modes of electrons. These collective modes 
correspond to the poles of the transverse
susceptibility, and  are obtained 
by summing up an infinite RPA series in the particle-hole channel with
the total momentum equal to either zero or $Q$.
The interaction vertices between fermionic quasiparticle and 
magnons have been 
calculated previously~\cite{CF}.  In the
strong coupling limit they are given by
\begin{eqnarray}
\Phi_{cc,vv} (k,q) &=& \left[\pm \Big(\epsilon^{(-)}_k+
\epsilon^{(-)}_{ k+q}\Big)\eta_q + \Big(\epsilon^{(-)}_k-
\epsilon^{(-)}_{k+q}\Big){\overline\eta}_q\right] \ , 
\nonumber \\
\Phi_{cv,vc} (k,q) &=& U~\left[
\eta_q \mp {\overline\eta}_q \right] \ .
\label{vertices}
\end{eqnarray}
where $\eta_q$ and ${\bar \eta}_q$ are given by 
\begin{equation}
\eta_q = \sqrt{S}~\left(\frac{1+\nu_q}{1-\nu_q}\right)^{1/4} \ ; \    
{\bar \eta}_q = \sqrt{S}~\left(\frac{1-\nu_q}{1+\nu_q}\right)^{1/4} \ ,
\end{equation}
and $\nu_q = (\cos q_x + \cos q_y)/2$.   

We see that there are two types of vertices: $\Phi_{cv,vc}$ 
which describes the interaction
between conduction and valence fermions, and 
$\Phi_{cc,vv}$ which involves either
only valence or only conduction fermions.
Apparently, the second diagram in Fig.~\ref{sigma} is more relevant since the  vertex which involves both conduction
and valence fermions scales as $U$. However, incident and intermediate fermions
in this diagram belong to different bands and are therefore
separated by a large, momentum
independent gap $\Delta \sim US$. As a result, the first diagram mostly 
contributes to the gap renormalization, which is exactly cancelled by
a renormalization of 
$\langle S_z \rangle $ such that the fully renormalized gap equals  
$2US$ as it indeed should be for the large $U$ Hubbard model \cite{CM1,CF}.
Expanding this diagram in $J/U$, we also obtain a momentum dependent 
term of $O(J)$ which contributes a
regular $1/S$ correction to the bare dispersion.

The first diagram in Fig.~\ref{sigma} involves only valence fermions. 
Here the vertex is reduced from $U$ due to the coherence factors and 
scales as $t$. At the same time, both incident and internal quasiparticles
are only $O(J)$ away from the Fermi surface which implies that the denominator
scales as $J$. The total contribution from the second diagram then behaves as
$JS(t/J\sqrt{S})^2$ and in addition is strongly 
momentum dependent.  Since the bare dispersion is of order $JS$,
the relative self-energy correction from the second diagram scales 
as $(t/J\sqrt{S})^2$ and is small only for extremely large
$S$. For physically relevant values of the spin, the expansion parameter
is obviously large, and one certainly cannot restrict with the second 
order in perturbation theory.

We now formulate precisely under which 
conditions we carry out the calculations. We 
assume that $S \gg 1$ and neglect all regular corrections in $1/S$. 
At the same time, we
assume that $t/J\sqrt{S} \gg 1$ and sum up an infinite series of diagrams 
in this parameter.
The restriction to large $S$ allows us not only to neglect the
 self-energy diagrams which
involve both valence and conduction fermions, but also to neglect 
the quantum corrections to the
spin-wave propagator. At half-filling, these regular $1/S$ 
corrections can,
with good accuracy, be absorbed into
the renormalization of the
 hopping term and the exchange interaction which are both input parameters for our calculations.

\begin{figure}  [t]
\begin{center}
\leavevmode
\epsfxsize=5cm
\epsffile{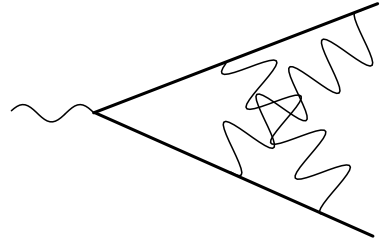}
\end{center}
\caption{ The lowest order vertex correction for the vertex between
fermions and transverse spin fluctuations.
 The diagram with only one wavy line is absent in the ordered state 
as it does not conserve the spin.}
\label{ver}
\end{figure}

The next step is to select the series of diagrams which have to be summed up.
To lowest order in perturbation theory, both 
self-energy and vertex corrections are equally relevant:
the self-energy correction yields
a relative contribution of $({\bar t}/J\sqrt{S})^2$,
while the leading order vertex correction shown in Fig.~\ref{ver} 
yields a relative factor 
$({\bar t}/J\sqrt{S})^4$ which is even larger. 
This result, however, 
changes if we estimate the strength of the self-energy and vertex corrections 
in a self-consistent manner, i.e.,
by considering all internal Green's functions and all vertices in the diagrams
in Figs.~\ref{sigma} and \ref{ver} 
as  full ones. This in turn yields self-consistent equations for the full
self-energy and the full vertex. Our self-consistent 
calculation of the self-energy 
correction is similar to the one performed by Kane, Lee and Read
(KLR)~\cite{KaneLee}.
Following KLR, we assume that the dominant pole approximation for the 
full fermionic Green's function is valid
upto energies of the order of the typical spin wave energy, i.e.,
the full Green's function can be approximated as 
$Z/(\omega - E_k)$ where $E_k = O(JS)$ (we later confirm this result
by explicit calculations). Substituting this form into the 
self-energy term and performing
standard manipulations we obtain for ${\bar t}/J\sqrt{S} \gg 1$ the
self-consistency condition
${\bar t}^2 Z^2 \Phi/J^2 S \sim 1$, where 
$\Phi$ stands for the vertex renormalization. It is
essential that there is only one power of $\Phi$ in this relation 
as only one of the two vertices in the self-energy diagram 
gets renormalized. 
On the other hand, in the vertex correction diagram, 
all vertices should be considered as full ones, and the self-consistency 
condition yields 
$(t^2 Z^2 \Phi/J^2 S)^2 \Phi^2 \sim 1$. Comparing these
two conditions, we obtain $Z \sim J \sqrt{S}/t$ and $\Phi = O(1)$. 
The result for $Z$ is
consistent with the one obtained by KLR. Clearly then, the self-energy
corrections are more relevant than the vertex corrections since the former 
reduce the quasiparticle residue to a parametrically small value, while the 
latter only change the vertex by a factor of order $O(1)$. 
Though the vertex corrections do not contain a factor $1/S$, it
 seems reasonable to assume that they just change
 the overall amplitude of the vertex but do not introduce 
any new physics. We therefore first neglect all
vertex corrections and obtain the full self-energy and thus the full Green's function in the self-consistent Born approximation~\cite{Schmitt}.
We then use the solution for the full Green's function
to estimate the relative strength of the vertex corrections. 
We find that the
vertex corrections change the vertex by roughly 
$20\%$ and therefore can be neglected with reasonable accuracy.

\begin{figure} [t]
\begin{center}
\leavevmode
\epsfxsize=7.5cm
\epsffile{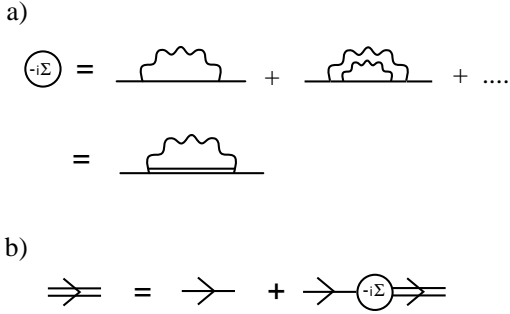}
\end{center}
\caption{$a)$ The self-energy is given by an infinite sum of  ``non-crossing"  diagrams. $b)$ The Dyson 
equation which together with the self-energy in $a)$ yields Eq.(\protect\ref{sc}).}
\label{rainbow}
\end{figure}

In the Born approximation, the full self-energy is  diagrammatically given by an infinite series of ``non-crossing" diagrams (see Fig.~\ref{rainbow}a).
Summing up this series, we obtain that the full self-energy 
has the same form as in second-order
perturbation theory, but the Green's function for the intermediate fermion 
is now replaced by the full one. 
The full Green's function is then obtained from  the Dyson equation (see Fig.~\ref{rainbow}b) and is analytically given by 
\begin{eqnarray}
&& G^{-1}(k,\omega) = \omega-(E_k^v - \mu) -\nonumber \\
&&\int{ \frac{d^2q}{4\pi^2}~d\Omega~\Psi(k,q) G(k+q,\omega+\Omega) F(q,\Omega)} \ ,
\label{sc}
\end{eqnarray}
where  $F(q,\omega)$ is the spin-wave propagator, and
\begin{eqnarray*}
\Psi(k,q) &=& \Phi_{vv}^2 = 32 S \bar{t}^2\Big[\nu_k^2+\nu_{k+q}^2-2\nu_k \nu_q \nu_{k+q} \\
&+& \sqrt{1-\nu_q^2}(\nu^2_{k+q}-\nu^2_k)\Big]/\sqrt{1-\nu_q^2} \ .
\end{eqnarray*}
The integration over the magnon momentum runs over the whole MBZ.

Eq.(\ref{sc})  is similar to the one derived earlier for the $t-J$ model
\cite{KaneLee,Marsi,Schmitt} with the only difference that Eq.\ref{sc}
contains the bare dispersion $E^v_k$. 
This dispersion is indeed also 
present when one derives the $t-J$ model from the
Hubbard model at  large $U$. However, it is due to  the three-site 
term which is usually omitted in the effective $t-J$ Hamiltonian~\cite{Lema}.

As we discussed in the introduction, the quasiparticle spectral weight
of the short-wavelength magnons in the parent compounds of the high-$T_c$ 
materials is likely to be strongly 
reduced as demonstrated by Raman and neutron scattering experiments. 
To account for this effect, 
we adopt a semi-phenomenological approach and 
introduce a cutoff,  $q_c$, in the integration over magnon 
momenta in the r.h.s.~of Eq.(\ref{sc}). 
We assume that for $q > q_c$, the magnon  
 spectral weight disappears into a broad background, and 
neglect the contribution 
to the self-energy from these $q$.
On the other hand,  for $q<q_c$, we 
 assume that the magnons are just free particles. 
Furthermore, for our analytical considerations, we will assume that 
$q_c$ is rather small such that we can
expand the dispersion of fermions and the spin-fermion vertex to linear order
in the magnon momentum. This last assumption is not well justified
as the magnitude of $q_c$ is unknown. 
Notice, however, that expanding upto leading order in $q_c$,
we  obtain two equal effective masses which are universal and 
independent of $q_c$. The smallness
of $q_c$ is then only needed for the  
corrections to these universal results to be small.
            
For free spin waves, we have 
\begin{equation}
F(q,\Omega) = {1 \over \Omega - \omega_q + i \delta} \ ,
\end{equation}
where $\omega_q =4JS\sqrt{1 - \nu^2_q}$ is the spin-wave spectrum. 
The magnon propagator has a pole in the lower half-plane of $\Omega$. 
In this half-plane, the
mean-field fermionic Green's function
$G(k,\omega) = (\omega - (E^v_k - \mu) + i\delta sgn \omega)^{-1}$
is free from nonanalyticities since $E^v_k - \mu <0$. 
We assume, following KLR, that the full $G(k,\omega)$ is also analytic
in the lower half-plane of
$\Omega$. Then one can straightforwardly 
perform the integration  
over magnon frequency in Eq.(\ref{sc}) and obtain
\begin{eqnarray}
G^{-1}(k,\omega)&=& \omega-(E_k^v -\mu) \nonumber \\
&& -\int{  \frac{d^2q}{4\pi^2}~\Psi(k,q) G(k+q,\omega+\omega_q)} \ .
\label{sc2}
\end{eqnarray}    
We first present our analytical results for the full Green's function
 in certain limiting cases, and then
 present the full numerical solution of Eq.(\ref{sc2}).

\section{Analytical Results}
\label{anal_res}

We  obtain the analytical solution of Eq.(\ref{sc2}) in two different 
ranges of $\omega$. In Sec.~\ref{anal_inc} we first solve the self-consistency 
equation in the limit 
$|\omega-\omega_{max}| \gg \Lambda$, where $\omega_{max}$  is the 
highest frequency at which the full Green's function first acquires a
finite imaginary part, and $\Lambda = JS ({\bar t}/J\sqrt{S})^{1/3}$.
 We show that for $|\omega-\omega_{max}| \gg \Lambda$
the excitations are purely incoherent and extend over a region of $\sim 
6{\bar t}\sqrt{2S}$.
In Sec.~\ref{anal_coh} we then consider the case $|\omega - \omega_{max}|
\leq \Lambda$. 
In this frequency range we find coherent excitations 
which exist up to energies of $O(JS)$ down from the maximal frequency. 

\subsection{Incoherent part of the excitation spectrum}
\label{anal_inc}
We first observe that the
 interaction vertex in Eq.(\ref{sc2}) has an overall scale of 
$({\bar t}\sqrt{S})^2$, 
while the
quasiparticle Green's function behaves as $1/\omega$ at very large frequencies
(here, and in the following, we shifted the frequency by the mean-field 
chemical potential, $\mu=-\Delta$). Obviously than, for $\omega \gg t\sqrt{S}$,
 the perturbative expansion in the spin-fermion
interaction is convergent, 
 and the density of states (DOS) is exactly equal to
zero, as in the mean-field theory. When $\omega$ is reduced to the scale of
${\bar t}\sqrt{S}$, the lowest-order self-energy term $\sim {\bar t}^2
S/\omega$ becomes of the same magnitude
 as the frequency in the bare Green's function, i.e, the expansion
parameter is $O(1)$. We show that in this frequency range 
there exists a critical value of $\omega$ 
below which perturbation theory becomes non-convergent 
and there appears  a finite DOS.
It is essential 
that for small $J\sqrt{S}/{\bar t}$, the critical frequency
is still much larger than the magnon frequency such that
one can neglect $\omega_q$ and $E^v_k$ compared to
$\omega$ in the r.h.s. of Eq.(\ref{sc2}).
The self-consistency equation
then reduces to a conventional integral equation 
\begin{equation}
G^{-1}(k,\omega)= \omega
-\int{  \frac{d^2q}{4\pi^2}~\Psi(k,q-k) \, G(q,\omega) }
\label{sc_inc}
\end{equation}
in which the dependence on the external momentum is only present in the
interaction vertex. Furthermore, we
 assume that $\omega \sim {\bar t} \sqrt{S}$ is
larger than the total magnon bandwidth, including the incoherent part.
In this situation, 
the integration over $q$ runs over the whole MBZ.

Before we present the solution of Eq.(\ref{sc_inc}), it is instructive to 
consider 
a simplified version of this equation in which 
$\Psi (k,q-k)$, which is a smooth function of the fermionic momentum,
 is just substituted by some constant $\sim {\bar t}^2 S$.
The equation for the full $G(\omega)$ then reduces to
\begin{equation}
G^{-1} (\omega) = \omega - {\tilde{\bar t}}^2 S G(\omega) \ ,
\label{toy1}
\end{equation}
 where ${\tilde{\bar t}}/{\bar t}=O(1)$. 
Solving this algebraic equation, we obtain for positive $\omega$
\begin{equation}
G(\omega) = \frac{2}{\omega + \sqrt{\omega^2 - \omega_{max}^2}} \ ,
\label{toy1a}
\end{equation}
where $\omega_{max} = 2 {\tilde{\bar t}} \sqrt{S}$. We see that for $\omega >
\omega_{max}$, the Green's function is real. This is the frequency range where perturbation theory is valid. For $\omega < \omega_{max}$, however,
the expression under the square root is negative, and the solution possesses
a finite imaginary part which gives rise to a finite DOS. The total width of
the DOS is obviously $W = 2\omega_{max} = 4 {\tilde{\bar t}} \sqrt{S}$.

We now solve Eq.(\ref{sc_inc}) with the 
actual $\Psi (k, q-k)$. We introduce a new function 
$f_k (\omega)$ via
\begin{equation}
G^{-1}_k(\omega)=\omega~ f_k(\omega^2) \ .
\end{equation}
Substituting this into Eq.(\ref{sc_inc}), we obtain
\begin{equation}
f_k = 1- \alpha \int { d^2q \over 4 \pi^2 f_q } 
\Bigg[ \nu^2_q-\nu^2_k +    
{ \nu_k^2+\nu_q^2-2\nu_k \nu_q \nu_{q-k} \over 
\sqrt{1-\nu^2_{q-k} } } \Bigg]  \  ,
\end{equation}
where we defined $\alpha=32 \bar{t}^2 S /\omega^2$. 

The general solution of Eq.(\ref{sc_inc}) can be obtained by expanding in the
eigenfunctions of the $D_{4h}$ symmetry group of the square lattice.
The solution is in general 
rather cumbersome because the vertex contains a
$k-$dependent term in the denominator. However, it is easy to
verify that  the expression in the square brackets vanishes when 
 $\nu_{q-k} \rightarrow 1$. 
The dominant contribution to the r.h.s.~of Eq.(\ref{sc_inc}) then 
comes from 
the  region of $q$-space where $\nu_{q-k}$ is relatively small i.e., the
denominator is close to one. For simplicity, we just set
it equal to one. We then obtain
\begin{equation}
f_k = 1- \alpha \int { d^2 \over 4 \pi^2 f_{q} } \, \nu_{q} 
\, \Big[ \nu_{q}-\nu_k \, \nu_{q-k} \Big] \ .
\label{sc_inc2}
\end{equation}
This equation is much simpler to solve because the
decomposition of $\nu_{q-k}$ into the eigenfunctions of the 
square lattice involves only four eigenfunctions:
\begin{equation}
\nu_{k-q} = \nu_k \nu_q + {\tilde \nu}_k {\tilde \nu}_q +
{\bar\nu}_k {\bar \nu}_q + {\bar{\tilde \nu}}_k {\bar{\tilde \nu}}_q  \  , 
\end{equation} 
where 
\begin{eqnarray}
\nu_k &=& \frac{1}{2} (\cos k_x + \cos k_y) \ ;~~~~ 
\tilde{\nu}_k = \frac{1}{2} (\cos k_x - \cos k_y) \ ; \nonumber \\
\bar{\nu}_k &=& \frac{1}{2} (\sin k_x + \sin k_y)\ ;~~~~
\bar{\tilde{\nu}}_k = \frac{1}{2} (\sin k_x - \sin k_y) \ . 
\label{eigenf}
\end{eqnarray} 
We now choose a general ansatz for $f_k$ consistent with Eq.(\ref{sc_inc2})
\begin{equation}
f_k=A+B \, \nu_k^2 +C \, \nu_k \tilde{\nu}_k + D \, \nu_k \bar{\nu}_k
+ E \, \nu_k \bar{\tilde{\nu}}_k 
\label{fk}    
\end{equation}
and solve this set of self-consistent algebraic equations for the 
coefficients.
We found that the coefficients $C, D$ and $E$ are equal to zero, while
$A$ and $B$ are the solutions of two coupled equations
\begin{eqnarray}
A&=& 1-2 \alpha \int \frac{d^2 q}{4\pi^2} \, {\nu^2_{q} \over
A+B \,\nu^2_{q} }  \ , \nonumber \\
B&=& 2 \alpha \int \frac{d^2 q}{4\pi^2} {\nu^2_{q} \over
A+B \,\nu^2_{q} } \ . 
\end{eqnarray}
Introducing $A=1-2\alpha x, B=2\alpha x$ and separating 
real and imaginary 
parts of $x$ by introducing $x=x_1+ix_2$, 
we obtain an equivalent set of equations for $x_1$ and $x_2$ 
\begin{eqnarray}
x_1 &=& \int \frac{d^2 q}{4\pi^2}  {\nu_{q}^2 \, \Big(1-2 \alpha x_1 
(1-\nu_{q}^2) \Big) \over \Big[1-2 \alpha x_1 
(1-\nu_{q}^2) \Big]^2 + 4 \alpha^2 x_2^2 (1-\nu_{q}^2)^2 } \ , \nonumber \\
x_2&=&\int \frac{d^2 q}{4\pi^2} { \nu_{q}^2 \, (1-\nu_{q}^2) (2\alpha x_2)
\over \Big[1-2 \alpha x_1 (1-\nu_{q}^2) \Big]^2 + 
4 \alpha^2 x_2^2 (1-\nu_{q}^2)^2 } \ .
\label{x1x2}
\end{eqnarray}
In terms of $x_1$ and $x_2$, the quasiparticle Green's function is given by
\begin{equation}
G(k, \omega)= {1 \over \omega}~{ 1-2 \alpha x_1(1-\nu_{k}^2)
+ i 2 \alpha x_2 (1-\nu_{k}^2) \over \Big( 1-2 \alpha x_1(1-\nu_{k}^2)
\Big)^2 +  4 \alpha^2 x_2^2 (1-\nu_{k}^2)^2 } \ .
\label{g_inc}
\end{equation}
Obviously, the spectral function and hence the DOS are finite
when $x_2 \neq 0$. 

A simple analysis of Eq.(\ref{x1x2}) shows that the solution with 
$x_2=0$ exists only for
$|\omega| > \omega_{max} = 2.97 \bar{t} \sqrt{2S}$ (or $\alpha <
\alpha_{cr} = 0.448$). At the critical point, we obtain $x_1 = 0.43$.
For smaller frequencies Eq.(\ref{x1x2}) yields 
a solution with finite imaginary part, just as we found with the toy model
with momentum independent $\Psi$.
The total bandwidth is equal to 
$W = 2 \omega_{max} \approx 6 
\bar{t} \sqrt{2S}$ upto corrections of order $O(JS)$ which we neglected. 
For $\omega$ only
slightly below $\omega_{max}$, we have 
\begin{equation}
x_2 \sim \sqrt{\omega_{max}-\omega} \ . 
\label{aa}
\end{equation}
Substituting this into Eq.(\ref{g_inc}), we obtain
 that the DOS behaves near $\omega_{max}$ as
\begin{equation}
N(\omega) \sim {1 \over \bar{t} \sqrt{S}} \, \left(\omega_{max}-\omega  
\over \omega_{max} \right)^{1/2} \ .
\label{dos}
\end{equation}

The above results are valid only as long as one can neglect the 
magnon dispersion.
We now estimate the range of validity of this approximation. 
Recall that in transforming
Eq.(\ref{sc2}) into Eq.(\ref{sc_inc}), we omitted the term
\begin{equation}
\int \frac{d^2q}{4\pi^2}~\Psi(k,q-k) 
[G(q,\omega + \omega_q) - G(q,\omega)]  \ .
\label{DG}
\end{equation}
Far from $\omega_{max}$, we do not expect this term to be 
relevant. Near the
maximum frequency, $G(\omega) - G (\omega_{max}) \propto (\omega_{max} -
\omega)^{1/2}$, and $\partial G /\partial \omega$ is singular.  
Substituting the form of $G$
from Eq.(\ref{g_inc}) with $x_2$ from Eq.(\ref{aa})
into Eq.(\ref{DG}) we find that the 
term we omitted can be neglected when 
$|\omega_{\max} -\omega| \geq J^2 S^{5/2}{\bar t}/(\omega_{\max} -\omega)^2$, 
i.e., when $|\omega_{\max} -\omega| \geq \Lambda$ where $\Lambda =
JS ({\bar t}/J\sqrt{S})^{1/3}$.                          
At  frequencies closer to $\omega_{max}$, the magnon dispersion is not
negligible, and the calculation of the spectral function should be
done using the full self-consistency 
equation Eq.(\ref{sc2}).  We will proceed with this calculation in the next section.

\subsection{Coherent part of the excitation spectrum}
\label{anal_coh}

In this section, we study the form of the quasiparticle Green's function
close to the top of the valence band, i.e., in the region 
$|\omega_{max} - \omega| \leq \Lambda$.

It is again instructive to consider first a toy model with a momentum
independent interaction. Assume that typical value of the magnon frequency
is ${\tilde J}S$ with  ${\tilde J}/ J=O(1)$. We then have instead of Eq.(\ref{toy1})
\begin{equation}
G^{-1} (\omega) = \omega - {\tilde{\bar t}}^2 S G(\omega + {\tilde J}S) \ .
\label{toy11}
\end{equation}
In the vicinity of $\omega_{max}$, the solution of this equation is
\begin{equation}
G(\omega) \approx \frac{2}{\omega_{max}}~
\left(1 + \left(\frac{2}{\omega_{max}}\right)^{1/2}~
\frac{((\omega - \omega_{max})^3 + 
{\tilde\Lambda}^3)^{1/2}}{|\omega - \omega_{max}|}\right)
\label{toy2}
\end{equation}
where $\omega_{max} = 2 {\tilde{\bar t}} \sqrt{S} + O({\tilde J}S)$ 
and ${\tilde \Lambda} =
{\tilde J}S ({\tilde{\bar t}}/{\tilde J}\sqrt{S})^{1/3}$.                          
We see that there are two typical scales introduced by ${\tilde J}$. For 
$|\omega - \omega_{max}| \geq \Lambda$, $G(\omega)$ in Eq.(\ref{toy2}) differs
from that in Eq.(\ref{toy1a}) only by small corrections. For $JS \leq 
|\omega - \omega_{max}| \leq \Lambda$, the frequency dependence of the full
solution is different from that in Eq.(\ref{toy1a}), however, $G(\omega)$ 
remains approximately equal to $2/\omega_{max}$. Finally, at 
$|\omega - \omega_{max}| \leq {\tilde J}S$, the full Green's function begin to
increase, and very near $\omega_{max}$ we have
\begin{equation}
G(\omega) \approx \frac{{\tilde J} \sqrt{S}}{\tilde{\bar t}}~
\frac{1}{\omega - \omega_{max}} \ .
\label{toy3}
\end{equation}
We see that very near $\omega_{max}$, the Green's function has a conventional
pole with the residue $Z = {\tilde J} \sqrt{S}/{\tilde{\bar t}}$.  This implies that
around $\omega_{max}$, there should exist coherent fermionic excitations.
 
We now proceed with the solution of the actual self-consistency equation with a
momentum-dependent $\Psi (k,q-k)$. 
Inspired by the solution of the toy model,  we assume that there exists
a frequency, $\omega_{max}$  for which
 $G^{-1} (k,\omega_{max})
=0$ at some $k=k_0$, and 
which differs from the previously found onset frequency only by an amount 
of $O(JS)$. We will not be able to fully verify this assumption 
analytically as it
would require us to find a solution of Eq.(\ref{sc2})  for all 
$k$ and $\omega \sim
\omega_{max}$ which we cannot do. However, we will later verify this assumption
in our numerical studies. We also assume and then verify
that $ k_0=(\pi/2,\pi/2)$, and 
that near $k=k_0$ and $\omega =
\omega_{max}$, the excitations
 are mostly coherent, and  the
quasiparticle Green's function has the form
\begin{equation}
G(k,\omega) = {Z \over \omega - \omega_{max} +E_{ k}
-i \gamma (\omega - 
\omega_{max})^2 \Theta (\omega - \omega_{max})} \ .
\label{ans}
\end{equation}
Here $Z_k$ is the quasiparticle residue, $\gamma$ is the damping coefficient,
$\Theta (x) = 1 (0)$ if $x<0$ ($x>0$), and
 the hole excitation spectrum 
has the form
\begin{equation}
E_k =  { (k_\perp -k_0)^2 \over 2 m_\perp } +
 { (k_\parallel -k_0)^2 \over 2 m_\parallel} \ ,
\end{equation}
where $k_\perp,  k_\parallel$ are the 
momenta along the boundary of the MBZ and 
along the zone diagonal, respectively. 

In addition, as we discussed above, we introduce an 
upper cutoff $q_c \leq 1$ in the integration over the magnon momentum, 
and restrict with an expansion of the magnon energy upto linear 
order in $q$. We recall that 
physically, the presence of this cutoff reflects
the experimental fact that the 
zone-boundary magnons cease to exist as well-defined quasiparticles
 and therefore effectively do not contribute to the self-energy of 
the valence fermions. 
We will see
that the quasiparticle residue $Z$ scales as $(q_c)^{-1/2}$, 
but the effective masses are independent of $q_c$.     

We now substitute the 
coherent ansatz for $G(k,\omega)$ into the self-consistency equation
Eq.(\ref{sc2}). 
Expanding around $k_0$ and $\omega_{max}$ and using the fact that
$G^{-1} (k_0,\omega_{max}) =0$,
we obtain self-consistent solutions for the
quasiparticle residue, the quasiparticle spectrum and
the damping coefficient. Consider first the
quasiparticle residue. Setting $k=k_0$ and expanding the r.h.s.~of the
self-consistency equation Eq.(\ref{sc2}) to linear order in $\omega_{max} -
\omega$ we obtain
\begin{equation}
\frac{1-Z}{Z} = \int { d^2q \over 4\pi^2}   
\Psi(k_0, q) 
{Z\over (\omega_q + E_{k_0 +q })^2 } \ ,
\label{om_cr1}
\end{equation}  
where the integration runs upto $q_c$.
Since $q_c \ll 1$, we can expand the two terms in the denominator 
to linear order in $q$. As
$\omega_q \propto q$ and $E_{k_0 +q} \propto q^2$,
the first term is  dominant. Performing the integration with only $\omega_q$ in
the denominator, we obtain
\begin{equation}
1= Z + {\sqrt{2} \bar{t}^2 Z^2 q_c \over \pi J^2 S } \  .
\label{z1a}
\end{equation}
In the limit $J \sqrt{S}/t \ll 1$, the term linear in 
$Z$ can be neglected and we find
\begin{equation}
Z= {J \sqrt{S} \over  \bar{t} } \; \Big({\pi \sqrt{2} \over q_c}\Big)^{1/2} \ .  
\label{z1b}
\end{equation}
We see that $Z$ scales linearly with $J \sqrt{S}/t$ as in our toy model.
This dependence was also obtained in earlier studies 
\cite{KaneLee}. It was however 
noticed in Ref.~\cite{Marsi} that the linear dependence
exists only for very small $J/t$. These authors argued that 
for moderate $J/t$, $Z \sim (J/t)^{1/2}$. We also found deviations
from the linear behavior already for moderately small $J/t$, however, 
we did not find a square root dependence for intermediate $J/t$.
A plot of $Z$ versus $J/t$ is presented in Figs.~\ref{zk2} and \ref{zk1}.

Next, we calculate the quasiparticle damping coefficient $\gamma$.
For this we again set $k=k_0$, neglect $E_{k_0 + q}$ compared to
$\omega_q$, but do not expand in $\omega - \omega_{max}$. However,
 since we are interested in small
deviations from $\omega_{max}$ we can 
neglect the damping term
on the r.h.s.~of Eq.(\ref{sc2}) compared to $\omega - \omega_{max}$.
The r.h.s.~of the self-consistency equation then takes the form
\begin{equation}
\int { d^2q \over 4\pi^2}   
\Psi(k_0, q) 
{Z\over \omega - \omega_{max} + \omega_q  + i\delta}  \ .
\label{om_cr2}
\end{equation}
Clearly, for $\omega >\omega_{max}$, the denominator is positive and
the integral does not contain an imaginary part. For $\omega < \omega_{max}$,
however, the integrand has a pole at $\omega = \omega_{max} - \omega_q$.
Integrating around the pole, we obtain a finite imaginary part which in 2D
scales as $(\omega- \omega_{max})^2$. After performing the explicit 
calculations, we obtain
\begin{equation}
\gamma= { \bar{t}^2 Z^2 \over (2S)^2 J^3 } \ .
\label{aaa}
\end{equation}
The same result was obtained earlier by Kane, Lee and Read~\cite{KaneLee}.
Note in passing that in contrast to a recent claim in Ref.~\cite{Mukhin}, 
we did not find a missing factor of 2 in their formula.
Substituting the expression for $Z$ into Eq.(\ref{aaa}), we finally obtain
\begin{equation}
\gamma = \frac{1}{4JS}~\frac{\pi \sqrt{2}}{q_c} \ .
\label{gam}
\end{equation}
Comparing now the damping term with the term linear in frequency, we find that
the fermionic excitations are weakly damped for $E_k = 
\omega_{max} - \omega \leq 4JS (q_c/\pi \sqrt{2})$.
For small $q_c$, this condition is satisfied
only in a small region around $k_0$. For example, for $q_c=\sqrt{\pi/2}$ and $m^{-1} \sim 4JS$, the fermionic excitations 
are only weakly damped for $|k-k_0| < 0.75$ which constitutes only a small
fraction of the MBZ. 
Away from this region, the
damping term is dominant, and the spectral function should possess a broad
maximum around $\omega=E_k$ rather than a sharp quasiparticle peak. 
This is in full agreement with the data~\cite{Wells,LaRosa}
 which show that the spectral 
function measured in photoemission experiments possesses a
clearly distinguishable
 quasiparticle peak only in the vicinity of $k_0$.

We now proceed with the calculation
of the effective masses. For this
we set $\omega = \omega_{max}$ and expand in the magnon
momentum. We restrict ourselves to the strong coupling 
limit $J \sqrt{S} \ll \bar{t}$ and neglect the bare dispersion, 
which in this limit is completely overshadowed by the self-energy correction. 
We then obtain
\begin{eqnarray}
{E_k \over Z } &=&  
- \int { d^2q \over 4\pi^2} \Psi(k_0, q) \nonumber \\
& & \hspace{-0.5cm} \times \Bigg\{ 
G(\omega_{max}+\omega_q,  k + q)- 
G(\omega_{max}+\omega_q,  k_0 +  q)
\Bigg\}  \nonumber \\
& & \hspace{-1cm} - \int { d^2q \over 4\pi^2} \Bigg\{ \Psi( k,  q) - 
\Psi(  k_0,  q) \Bigg\} 
G(\omega_{max}+\omega_q, k_0 + q) \nonumber \\
& & \hspace{1cm} - \int { d^2q \over 2\pi^2} \Bigg\{ \Psi( k,  q) - 
\Psi( k_0,  q) \Bigg\} \nonumber \\
&& \hspace{-1cm}\times \Bigg\{ 
G(\omega_{max}+\omega_q, k +  q)- G(\omega_{max}+\omega_q, 
k_0 + q) \ .
\Bigg\}
\label{ak2}
\end{eqnarray}
Expanding the quasiparticle Green's function, we obtain
\begin{eqnarray}
& & G(\omega_{max}+\omega_q,  k + q)- 
G(\omega_{max}+\omega_q, k_0 +  q) =
\nonumber \\
& & Z {E_{k_0 +  q} - E_{ k + q} \over 
(\omega_q + E_{ k_0 +  q})^2 } + 
Z {(E_{ k_0 +  q} - E_{ k +  q})^2 \over 
(\omega_q + E_{k_0 + q})^3 }  + ...
\end{eqnarray}
where 
\begin{equation} 
E_{k + q} - E_{k_0 +  q} = 
 { k^2_\perp \over 2 m_\perp } + { k^2_\parallel \over 2 m_\parallel} 
+ { k_\perp q_\perp \over m_\perp } + 
{ k_\parallel q_\parallel \over m_\parallel} \ . 
\label{exp_ak}
\end{equation}
The expansion of  $\Psi(k,  q)$  upto quadratic order
 in $ k$ and upto 
linear order in $ q$ yields
\begin{eqnarray}
\Psi(k, q) - \Psi( k_0,q) &=& 32\bar{t}^2 S \nonumber \\
&& \hspace{-1cm} \times \Bigg\{ {\sqrt{2} \over 4} k_\perp^2 q (1- { q_\perp^2 \over q } ) - k_\perp q_\perp - { k_\parallel^2 \; q_\perp^2 \over 2 \sqrt{2} q } \Bigg\} \ .
\label{exp_psi} 
\end{eqnarray}
Inserting now these expressions into 
Eq.(\ref{ak2}) and using the result for the quasiparticle residue,
we find that one 
of the two contributions to the first term on the r.h.s.~of Eq.(\ref{ak2})
cancels out the $E_k/Z$ term on the l.h.s. The remaining terms are all
proportional to $\Psi$, and therefore the energy scale $t$ drops completely 
out of the problem. The only remaining scale is given by $\omega_q$, and the inverse
effective masses are therefore proportional to the spin-wave velocity. 
Furthermore, we found that the integrals over the magnon momentum in 
the remaining terms in Eq.(\ref{ak2}) are
confined to the upper limit of the $q-$integration, and 
 all scale as $q^2_c$. 
Therefore, the cutoff $q_c$
 also drops out of the problem. As a result, we
obtain universal, model-independent equations for the masses
\begin{eqnarray}
{ 3\over (4 J S m_{\perp})^2} -  { 1\over J S m_\perp} + 1 &=& 0 \ , \nonumber \\
 { 1 \over (4 J S m_\parallel)^2} -1 &=& 0 \  .
\label{sc_mass}
\end{eqnarray}
The second equation yields $m_\parallel^{-1} = 4JS$, 
while for $m_{\perp}$ we obtain two solutions:
$m_{\perp}^{-1} =  4JS \ \mbox{or} \ { 4 \over 3} JS$. 
We have checked that only the first solution for $m_{\perp}$
can be continuously connected with the perturbative solution at weak coupling.
 Then only the
first solution is physically relevant, and we finally obtain
\begin{equation}
m_{\parallel} = m_{\perp} = (4JS)^{-1}\ .
\label{masses}
\end{equation}
We see that in the limit $q_c \ll 1$ and for $J \sqrt{S} /t \ll 1$ when 
the bare dispersion can be 
neglected, the effective masses are equal, i.e., the top of the
valence band is isotropic. This result is an intrinsic 
property of the Hubbard (or $t-J$) model at strong coupling,
independent of the form of the bare hopping.
Note that the value of the masses is exactly
the same as in the mean-field theory with $t^{\prime} = -0.5J$.

\begin{figure} [t]
\begin{center}
\leavevmode
\epsfxsize=7.5cm
\epsffile{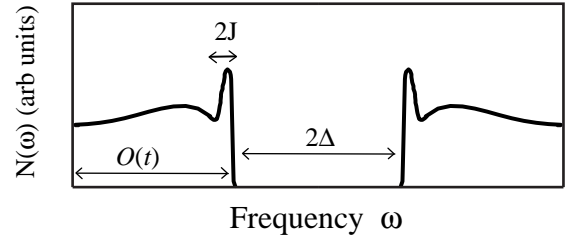}
\end{center}
\caption{The schematic form of the DOS at half-filling. 
The DOS reaches a maximum
at energies $\sim JS$, below the gap, then drops down at slightly 
larger energies, and then gradually increases and
saturates at energies which are $\sim t$ away from the gap.}
\label{DOS}
\end{figure}

We also estimated the magnitude of the corrections to this universal result for the masses.  We indeed found that 
as $q_c$ increases, the dispersion becomes
anisotropic with $m_{\perp} > m_{\parallel}$. This trend is consistent with the
results of other authors who integrated over the full magnetic Brillouin zone
in Eq.(\ref{sc2}) \cite{Marsi,Dag2,Manous}.
Finally, the form of the coherent part of the Green's function  
in Eq.(\ref{ans}) implies that the DOS behaves
 as $(\omega_{max} - \omega)^2$ very near $\omega_{max}$ and reaches the value
\begin{equation}
N \sim {Z \over JS} \sim {1 \over t \sqrt{S} }
\end{equation}
at $\omega_{max}-\omega \sim JS$, which is the largest scale where 
this form is applicable. At even larger frequencies, the DOS scales as
\begin{equation}
N \sim {Z \over \omega_{max}-\omega}
\end{equation}
and transforms into the fully incoherent DOS given by Eq.(\ref{dos}) at
$\omega_{max}-\omega \sim \Lambda$. This incoherent density of states then gradually increases with frequency and saturates at
$\omega \ll \omega_{max}$. These  results imply
 that the DOS reaches a maximum
at $\omega_{max} - \omega \sim JS$, then drops down at slightly 
larger frequencies $\sim \Lambda$, and then gradually increases and
passes through a broad maximum  at $\omega \ll \omega_{max}$.
The behavior of the DOS is presented in Fig.~\ref{DOS}.
This behavior is in agreement with the numerical results which also
find a strong coherent peak in the DOS at $\omega_{\max} - \omega \sim J$
 on top of a smooth incoherent background~\cite{Preuss}.

\subsection{Vertex corrections}

Finally, we consider the effect of vertex corrections. We already 
found in Sec.~\ref{form} that these corrections do not introduce
 any new scale, but 
at the same time they do not possess a factor of $1/S$ and therefore can 
only be
neglected due to a numerical smallness. To estimate the magnitude of the
vertex renormalization, we computed the lowest-order vertex correction
shown in Fig.~\ref{ver} with the  full quasiparticle Green's functions 
from Eq.(\ref{ans}).
We followed the same computational steps as
before, i.e., expanded to linear order in the magnon momentum and
integrated upto $q_c$. Performing these calculations, we obtain
that at small
external momenta 
the full vertex has the same functional form  as the bare one and  
differs from it by a factor $(1 + \delta)$
where 
\begin{equation}
\delta = \left(\frac{\sqrt{2} {\bar t}^2 Z^2 q_c}{2\pi J^2 S}\right)^2~I
\label{vertex}
\end{equation}
and
\begin{equation}
I = \int_0^1 dx \int_0^1 dy~\frac{x y}{(x+y)^2} =
(\log 2 - 0.5) \approx 0.2 \ .
\end{equation}
Substituting  Eq.(\ref{z1b}) for $Z$ into Eq.(\ref{vertex}),
we obtain that the term in brackets is equal to unity, i.e., $\delta =
I \approx 0.2$. We see that the leading vertex correction accounts 
for only a $20\%$ renormalization of the bare vertex. We did not explicitly 
compute higher-order vertex corrections, but
our estimates show that they are likely to be progressively smaller. 
We therefore estimate that 
our analytical and numerical results for the dispersion
obtained without vertex corrections are valid with an accuracy of about 
$20\%$.

\section{Numerical Results}
\label{num_res}

We now proceed with the discussion of the 
full numerical solution of the self-consistency 
equation for the quasiparticle Green's function. 

As in the previous section, we begin by considering
in Sec.~\ref{num_inc} the frequency range $|\omega-\omega_{max}| \gg \Lambda$, 
in which the spectrum is completely incoherent. 
In Sec.~\ref{num_coh} we then consider frequencies close to $\omega_{max}$ for 
which we obtain coherent excitations on the scale of $O(JS)$.
We  present the results for the dispersion of a single hole
for different values of $J/t$ and $t^{\prime}/t$, as well as different cutoffs
 $q_c$. For comparison with earlier studies we also present the results for
 the case when the magnons are considered as free particles. 

We will demonstrate that
for $t^{\prime} =0$, one recovers two equal effective masses only for 
very large $t/J$. However, after including a nearest-neighbor hopping  $t^{\prime} =-0.5J$, we obtain two roughly equal masses for all values of 
$t/J$. In this situation,  the only effect of the decrease of $J/t$ 
is the transfer
of the spectral weight from the coherent to the incoherent part
of the dispersion. Furthermore, for the experimentally relevant case $J/t=0.4$, we find $m^{-1} \approx (4JS)^{-1}$. For $S=1/2$ this yields
$m^{-1} \approx (2J)^{-1}$, which is the same value that was obtained 
in the photoemission experiments on $Sr_2CuO_2Cl_2$.

\subsection{Incoherent part of the excitation spectrum}
\label{num_inc}  

As we discussed in Sec.~\ref{anal_inc},
 for $|\omega-\omega_{max}| \gg \Lambda$ 
we can neglect the magnon dispersion on the r.h.s~of Eq.(\ref{sc}) and
consider an integral equation only in momentum space. Following the same 
argument we also neglected the bare fermionic dispersion  in Sec.~\ref{anal_inc}. For our numerical studies, however, we kept the bare fermionic
dispersion in order to illustrate how the DOS evolves with  $J/t$.
\begin{figure} [t]
\begin{center}
\leavevmode
\epsfxsize=7.5cm
\epsffile{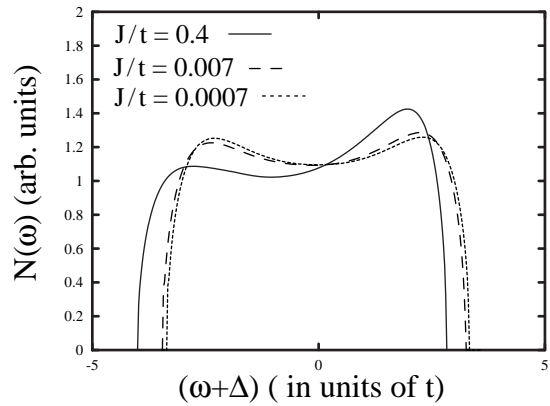}
\end{center}
\caption{The incoherent part of the hole excitation spectrum for several
values of $J/t$  
(solid line $J/t=0.4$, dashed line $J/t=0.007$ and dotted lines $J/t=0.0007$).}
\label{incoha}
\end{figure}
In Fig.~\ref{incoha} we present for several values of $J/t$ 
the DOS resulting from the numerical solution 
of Eq.(\ref{sc_inc}) for $S=1/2$ and $t^{\prime} =0$.
We see that for intermediate $J/t =0.4$ (solid line), 
the DOS is asymmetric around $\omega =0$ with the
density shifted towards negative frequencies. This indicates that 
the contribution from the bare
dispersion which by itself yields a finite DOS 
only for negative $\omega$ is not negligible.
 With decreasing $J/t$ the asymmetry becomes weaker,
until it basically vanishes for $J/t=0.007$. 
This result is expected since in the limit $J/t \rightarrow 0$ the bare 
dispersion becomes irrelevant, and we should
recover a symmetric DOS.
We also found that the total bandwidth only weakly depends on $J/t$ and
is roughly equal to $W=6.6 t $. This value is only
 slightly larger than $W=6 t$
 which we obtained analytically in Sec.~\ref{anal_inc}.

\subsection{Coherent part of the excitation spectrum}
\label{num_coh}

In order to solve Eq.(\ref{sc}) for the full Green's function
we use a discrete mesh in frequency and in $k-$space. 
We assume that near the top of the valence band
the quasiparticle Green's function has the form presented 
in Eq.(\ref{ans}) and obtain the onset frequency 
$\omega_{max}$  
and the hole dispersion $E_k$ from the conditions 
\begin{eqnarray}
G^{-1}(k=k_0,\omega=\omega_{max}) &=& 0  \ ,\nonumber \\
G^{-1}(k,\omega=\omega_{max}) &=& E_k/ Z \ .
\label{hole_disp}
\end{eqnarray}
To obtain the quasiparticle residue, we compute
$G^{-1} (k=k_0,\omega)$ and use the relation
\begin{equation}
Z = {\Delta \omega \over G^{-1}(k,\omega_{max}+\Delta \omega)-
G^{-1}(k,\omega_{max}) } \ , 
\end{equation}
where $\Delta \omega$ is a small shift from the maximal frequency.
The dispersion extracted from Eq.(\ref{hole_disp}) is formally valid only
in the vicinity of $k_0$. At larger distances from $k_0$, 
$E_k$ does not necessary coincide with
the position of the maximum in the spectral function due to a strong
quasiparticle damping. 
In our numerical procedure for 
solving the self-consistency equation, 
we relate the Green's function
at a given frequency $\omega$ to the Green's functions at larger
$\omega + \omega_q$,
and progressively compute $G$ at smaller and smaller $\omega$. Using this
method, we cannot obtain the imaginary part 
of the full Green's function and therefore
are unable to compare 
 $E_k$ extracted from Eq.(\ref{hole_disp}) with the position of the maximum in
the spectral function. We just
 assume without proof that at least not too far from $k_0$, 
$E_k$ and the peak position roughly coincide.

\begin{figure} [t]
\begin{center}
\leavevmode
\epsfxsize=7.5cm
\epsffile{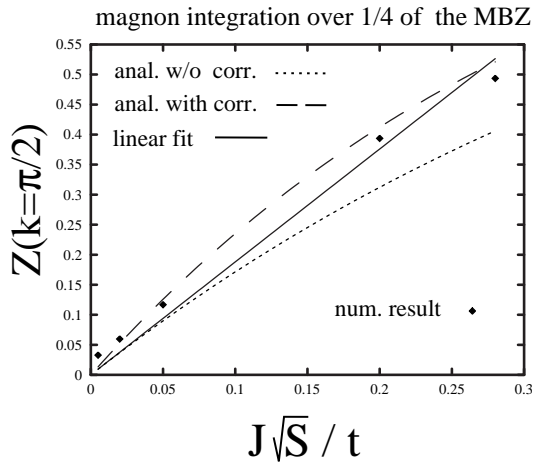}
\end{center}
\caption{$Z$ as a function of $J/t$. 
The dotted and solid lines are the plots of Eqs.(\protect\ref{z1a}) and 
(\protect\ref{z1b}), respectively.
The dashed line is our analytical result with subleading
corrections in $q_c$. The filled diamands are our numerical results. The 
integration over magnon momenta is restricted to $1/4$ of the MBZ.}
\label{zk2}
\end{figure}

\begin{figure} [t]
\begin{center}
\leavevmode
\epsfxsize=7.5cm
\epsffile{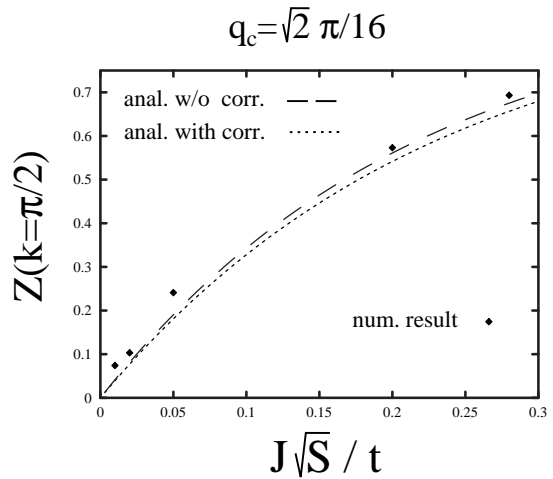}
\end{center}
\caption{The same as in Fig.~\protect\ref{zk2}  for  $q_c=2^{1/2}\pi/16$.}
\label{zk1}
\end{figure}

We first present in Figs.~\ref{zk2} and \ref{zk1} 
our results for the quasiparticle 
residue at $k=k_0 = (\pi/2, \pi/2)$ 
as a function of $J/t$. 
We have chosen two values of $q_c$: a smaller one
$q_c = \sqrt{2}\pi/16$ and a larger one for which 
the integration over the magnon momenta runs over $1/4$ of the MBZ.
The squares in these figures represent 
our numerical results, the dotted line
is our analytical formula, Eq.(\ref{z1b}),
 obtained to leading order in $q_c$, and the dashed line 
incorporates subleading corrections in $q_c$. 
We see that for smaller $q_c$, the agreement 
between the numerical data and the results to leading order in 
$q_c$ is rather good. For larger $q_c$, subleading
 corrections are more relevant. In both cases, however,
 the quasiparticle residue is
substantially reduced from its value $Z=1$ for free fermions
 already for moderate $J/t$. We also see that the linear dependence
exists only for very small $J/t$ (see linear fit in Fig.~\ref{zk2}). 

\begin{figure} [t]
\begin{center}
\leavevmode
\epsfxsize=7.5cm
\epsffile{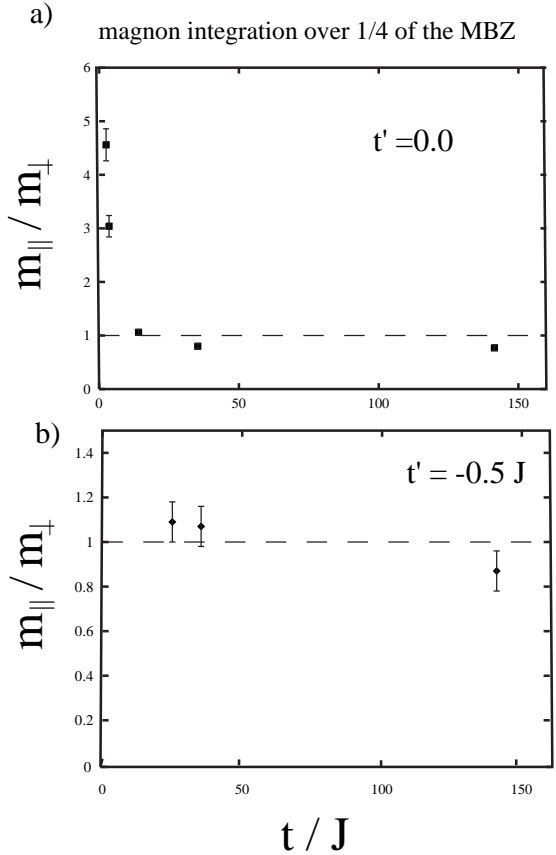}
\end{center}
\caption{The ratio 
of the effective masses $m_\perp$ and $m_\parallel$ 
as a function of $t/J$ for $(a)$ $t^\prime=0$ and $(b)$ $t^\prime=-0.5J$. 
The integration over the magnon 
momenta runs over $1/4$ of the MBZ.}
\label{massrati}
\end{figure}

\begin{figure} [t]
\begin{center}
\leavevmode
\epsfxsize=7.5cm
\epsffile{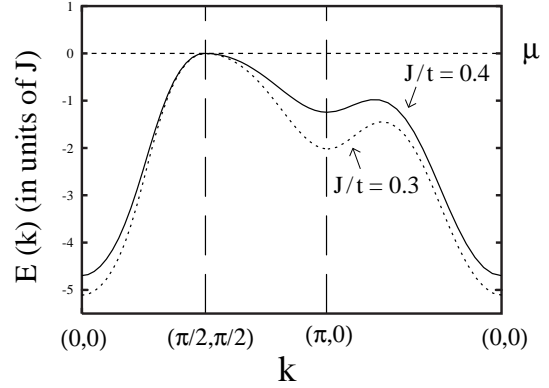}
\end{center} 
\caption{The fermionic dispersions for $t^\prime=0.0$ and two different 
values for $J/t$ (solid line $J/t=0.4$, dotted line $J/t=0.3$). 
The magnon integration is restricted to $1/4$ of the MBZ.}
\label{edisp2}
\end{figure}

\begin{figure} [t]
\begin{center}
\leavevmode
\epsfxsize=7.5cm
\epsffile{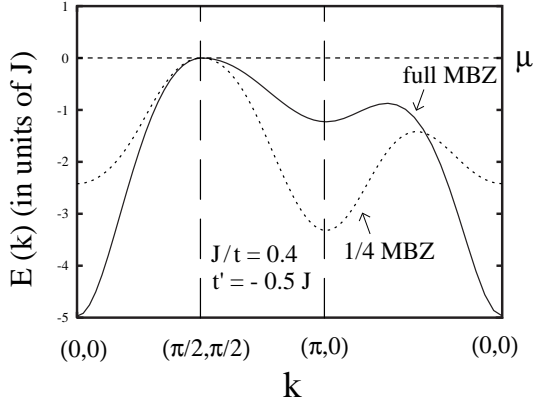}
\end{center}
\caption{The fermionic dispersions for $J/t=0.4$ and 
$t^\prime=-0.5J$ and two 
different ranges of integration. The solid and dotted lines 
are the results for the integration over the full MBZ and $1/4$ of the MBZ, 
respectively.}
\label{edisp6}
\end{figure}

\begin{figure} [t]
\begin{center}
\leavevmode
\epsfxsize=7.5cm
\epsffile{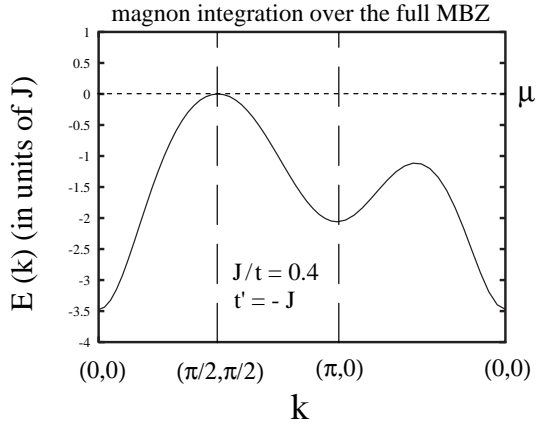}
\end{center}
\caption{The fermionic dispersions for $J/t=0.4$ and 
$t^\prime=-J$. The integration over magnon momenta extends over the full MBZ.}
\label{edisp7}
\end{figure}

In Fig.~\ref{massrati}  
we present the results for the ratio of the masses as a function of $t/J$
for $t^{\prime} =0$ and $t^{\prime} = -0.5J$, respectively. 
In both cases, the integration over the magnon 
momenta runs over $1/4$ of the MBZ. We see that for both values of 
$t^{\prime}$, the ratio of
the masses approaches one in the limit $t/J \rightarrow \infty$. This is in
full agreement with our analytical results. We also see, however, that for
$t^{\prime} =0$, one needs very large, unphysical values of $t/J$ to
recover the limiting behavior. For $t^{\prime}= -0.5J$, the ratio of the
masses is equal to one already at the mean-field level, and our results
demonstrate that the ratio remains roughly equal to one for all values of
$t/J$ including the experimentally relevant $t/J =2.5$. 
In order to see the effect of the $J/t$ ratio on the whole 
fermionic dispersion, we present the results for $E (k)$ 
for $t^{\prime} =0$ and two different values of $J/t$ in Fig.~\ref{edisp2}.
We clearly see that the variation of $J/t$ mainly
affects the dispersion around $(0,\pi)$.  
The excitation energy in this region increases 
with $t/J$, 
i.e., the dip in the dispersion becomes deeper,
which immediately leads to a decrease in the
ratio of the effective masses. At the same time, the
 overall bandwidth  only slightly increases with decreasing $J/t$.
\begin{figure} [t]
\begin{center}
\leavevmode
\epsfxsize=7.5cm
\epsffile{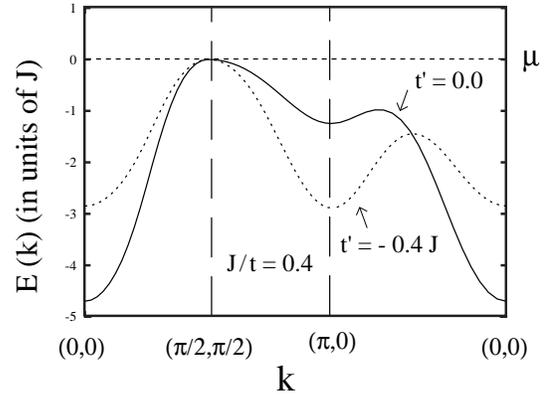}
\end{center}
\caption{The fermionic dispersions for $J/t=0.4$ and two different values 
for $t^\prime$ (solid line $t^\prime=0.0$, dotted line $t^\prime=-0.4J$). 
The magnon integration is restricted to $1/4$ of the MBZ.}
\label{edisp1}
\end{figure}
In Fig.~\ref{edisp1} we compare
 the results for the fermionic dispersion
 for $J/t =0.4$ and for two values of $t^{\prime}, t^{\prime} =0$
and $t^{\prime} =-0.4J$ 
(here the magnon integration runs over $1/4$ of the MBZ). 
We see that for $t^{\prime} =0$, the dispersion is rather anisotropic and
inconsistent with the experimental data~\cite{Wells,LaRosa}.
 On the contrary, for $t^{\prime} =-0.4J$, not
only the masses are equal, but also the energies at $(0,0)$ and $(0,\pi)$
are nearly equal to each other, 
and the bandwidth for coherent excitations is 
about $3J$. All three of these results are
in full agreement with the data~\cite{Wells,LaRosa}.
We also found that 
the energy at $(0,\pi/2)$ is about half of that at $(0,0)$ which agrees with
the most recent data by LaRosa {\it et al}~\cite{LaRosa}.
 The results for $t^{\prime} =-0.5J$ are
similar to those for $t^{\prime} =-0.4J$ and are presented in Fig.~\ref{edisp6}.
In this figure  we compare the dispersion
for $t^{\prime} =-0.5J$ for two different ranges of integration over the magnon momentum. 
We see that while the
integration over $1/4$ of the MBZ yields a dispersion 
roughly consistent with the data, the integration over the full 
MBZ yields a highly anisotropic dispersion.
However, one can increase $t^{\prime}$ even further and 
reduce  the energy at $(0,\pi)$ thus making the dispersion near
$(\pi/2,\pi/2)$ more isotropic even for the integration over the full MBZ.
We illustrate this in Fig.~\ref{edisp7} where we  present the results for the
fermionic dispersion  for $t^{\prime} =-J$ and for the integration over 
the full MBZ.
We see, however, that the overall bandwidth is still larger than 
in the experiments. We therefore conclude that if the integration over magnon
momentum runs over the full MBZ (which implies that magnons are treated as free
particles), the dispersion is inconsistent with the data for all reasonable 
values of $t^{\prime}$. In this situation, to account for the data
one has to  adjust the hopping to even further neighbors.

\begin{figure} [t]
\begin{center}
\leavevmode
\epsfxsize=7.5cm
\epsffile{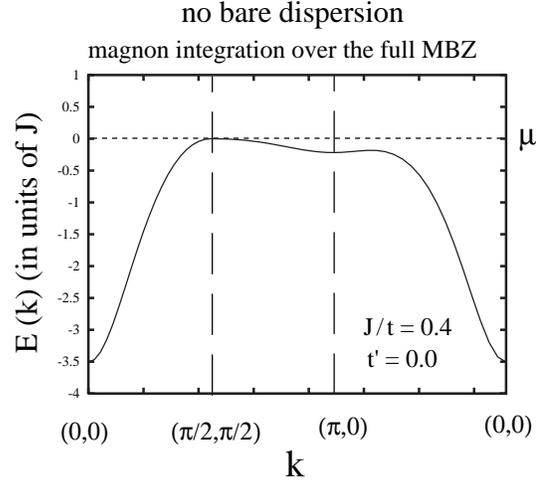}
\end{center}
\caption{The fermionic dispersion in the $t-J$ model for $J/t=0.4$
and $t^\prime=0$. The integration over magnon momenta runs over the 
full MBZ.}
\label{edisp4}   
\end{figure}

\begin{figure} [t]
\begin{center}
\leavevmode
\epsfxsize=7.5cm
\epsffile{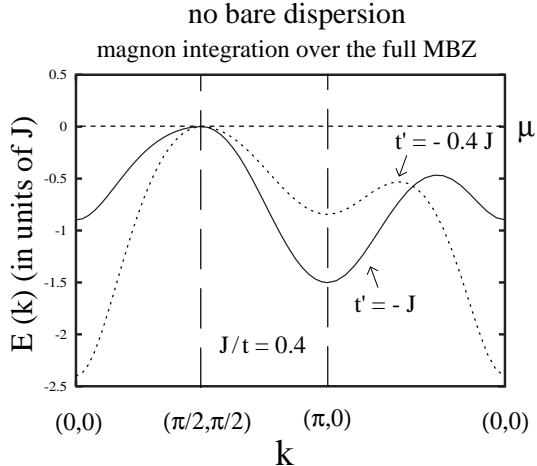}
\end{center}
\caption{The fermionic dispersion within the $t-J$ model for  
$J/t = 0.4$ and for two different values of $t^\prime$
 (dotted line $t^\prime=0.4J$, solid line 
$t^\prime=-J$). The integration over magnon momenta runs over the full MBZ.}
\label{edisp8} 
\end{figure}

For completeness, we also present several results for 
the integration over the
full MBZ in the conventional $t-J$ model without the three-cite term.
This corresponds to neglecting the bare fermionic dispersion in Eq.(\ref{sc}). In Fig.~\ref{edisp4} we present
the results for the excitation energy for $t^{\prime} =0$ and $J/t =0.4$.
This form of the dispersion 
is in very good agreement with the results of
earlier studies~\cite{Marsi,Dag2,Manous}.  As in previous studies, we found
that the effective mass along the zone diagonal is roughly $7$ times smaller
than the mass along the boundary of the MBZ.
In Fig.~\ref{edisp8}  we present the results for the evolution of the
dispersion with $t^{\prime}$.
We  see that as $t^\prime$ increases,
the effective mass along
$(0,\pi)$ direction gets smaller. However,
a rather large $|t^{\prime}|$ is needed to reproduce two equal masses.
Moreover, for equal masses, the overall bandwidth is about two times smaller
than in the experiments.  We see again that without restricting the integration
over magnon momentum, one needs to add and 
to fine tune the hopping parameters to even 
further neighbors to reproduce the experimental data.

\section{Summary}
\label{concl}

We now summarize the results of our studies. 
We considered in this paper the
dispersion of a single hole injected into a quantum antiferromagnet. We 
applied a spin-density-wave formalism extended to large 
number of orbitals $n = 2S$, 
and obtained an integral equation for the
full quasiparticle Green's function in the self-consistent ``non-crossing" 
Born approximation. 
At $S =\infty$, the mean-field theory is exact. At finite $S$, we found that
the self-energy correction to the mean-field formula for $G(k,\omega)$ scales 
as ${\bar t}/J\sqrt{S}$, and for large ${\bar t}/J$, 
relevant to experiments, is small only in the
unphysical limit of a very large spin. 
We found that for  ${\bar t}/J\sqrt{S} \gg 1$, the
bare fermionic dispersion is completely overshadowed by the self-energy
corrections. In this case, the quasiparticle Green's function contains a
broad incoherent continuum which extends over a frequency range of $\sim 6t$. 
In addition, there exists a narrow region
of width $O(JS)$ below the top of the valence band, where the excitations
are mostly coherent, though with a small quasiparticle residue $Z \sim
J\sqrt{S}/{\bar t}$. The top of the valence band is located at $(\pi/2,\pi/2)$.

We found that the form of the fermionic dispersion, and, in particular, the
ratio of the effective masses near $(\pi/2,\pi/2)$ strongly depend on
the assumptions one makes for the form of the magnon propagator. 
For free magnons, the
integration over magnon momenta in the self-energy  runs over the 
whole MBZ. In this case, we found, in agreement with earlier
studies, that the dispersion around $(\pi/2,\pi/2)$ is anisotropic with a 
much smaller
mass along the zone diagonal. This result holds even if the bare dispersion
contains a sizable $t^{\prime}$ term. 

We, however, argued in the paper that the two-magnon Raman 
scattering~\cite{Blum} as well as neutron scattering experiments~\cite{gabe}
 strongly suggest that
 the zone boundary magnons are not free particles since a substantial 
portion of
their spectral weight is transformed into an incoherent background. Most
probably, this transformation is due to a strong magnon-phonon interaction.
In this situation, only
magnons with small momenta, for which the interaction with phonons is necessary 
small, actually contribute to the self-energy.
We modeled this effect by introducing a cutoff $q_c$ in the integration over
 magnon momenta. 
We found analytically that for small $q_c$, 
the strong coupling
solution for the Green's function is universal, and both of the
effective masses are equal to $(4JS)^{-1}$.
We further studied numerically the shape
of the dispersion for intermediate $J/t$ and found that for 
$t^{\prime} \sim -0.5J$ the ratio of the
masses remains roughly equal to one for basically all values of $J/t$.
This particular value for $t^\prime$ is 
obtained from a comparison of the low energy excitations in the 
underlying three-band model and the effective one-band Hubbard
model~\cite{Hyber}. 
We computed the full fermionic
dispersion for $J/t =0.4$ relevant for $Sr_2CuO_2Cl_2$, 
and $t^\prime=-0.4J$ and
found that not only the masses are both equal to $(2J)^{-1}$, but also
the energies at $(0,0)$ and $(0,\pi)$ are 
equal, the energy at $(0,\pi/2)$ is about half of that at $(0,0)$,
and the bandwidth for the coherent excitations is around $3J$.
All of these results are  in full agreement with the experimental data
by LaRosa {\it et al}~\cite{LaRosa} and also 
 by Wells {\it et al}~\cite{Wells} (except for  a  slightly smaller
bandwidth and larger energy at $(0,\pi)$)
Finally, we computed the damping of the coherent fermionic 
excitations and found that 
it is small only in a narrow range around $(\pi/2,\pi/2)$.
Away from the vicinity of $(\pi/2,\pi/2)$, the excitations are overdamped,
and the spectral function possesses a broad maximum rather than a 
sharp quasiparticle peak.
This last feature was also reported in the photoemission experiments. 

One of the goals of the present paper was to demonstrate that the experimental
data for $Sr_2CuO_2Cl_2$ can be described without introducing a spin-charge
separation. In this respect, we predict that the data should not change much 
if the experiments are performed well below $T_c$ though some anisotropy
of the masses is possible because the spin damping decreases with decreasing
$T$ and hence $q_c$ should become larger. 
This prediction is contrary to the one derived
 from a model with spin-charge separation~\cite{Laughlin}. In this last case,
it was suggested that the minimal model with $t^{\prime} =0$ already
accounts for the key experimental features, and that well below $T_c$, 
spinons and holons are confined such that one
should recover a strong anisotropy of the masses, similar to that in 
Fig.~\ref{edisp4}    

A final remark.
Though the point of departure of our analysis is very different from the 
one in the scenario  based on spin-charge separation~\cite{Laughlin},
in many respects there exists a striking similarity between the results 
obtained in both approaches.
First, we found that the excitations are mostly incoherent, and the
bandwidth of incoherent excitations is several $t$. Second, we obtained that
the dispersing excitations 
observed in photoemission measurements exists upto an energy
scale which is given by  $J$  rather than by $t$.
Both of these results are in full agreement with the results
 obtained by Laughlin 
in the framework of spin-charge separation. 
However, contrary to Laughlin, we did find a conventional Fermi-liquid 
pole in $G(k,\omega)$ near $(\pi/2,\pi/2)$. The quasiparticle
residue of the coherent excitations is small in the strong coupling limit and
vanishes when $J/t \rightarrow 0$. In view of these results,
we suspect that spinons and holons are actually confined  even above the 
Neel temperature, but the confinement is weak near $(\pi/2,\pi/2)$ and 
disappears when $J/t \rightarrow 0$.
A detailed study of this confinement is clearly called for.

It is our pleasure to thank G. Blumberg, 
E. Dagotto, R. Joynt, R. Laughlin, M. Onellion, Z-X. Shen and 
O. Sushkov for
useful discussions. The work was supported by NSF-DMR 9629839 and by
A.P. Sloan Fellowship (A. Ch).

\end{document}